\documentclass[a4paper,11pt]{article}

\usepackage{pos}
\usepackage{url}
\usepackage{bm}
\usepackage{braket}

\title{Nuclear force with LapH smearing}
\author*[a]{Takuya Sugiura}
\author[b,c]{Yutaro Akahoshi}
\author[d,c]{Tatsumi Aoyama}
\author[e]{Takahiro M. Doi}
\author[c,a]{Takumi Doi}

\affiliation[a]{RIKEN Interdisciplinary Theoretical and Mathematical Sciences Program (iTHEMS), \\
Saitama 351-0198, Japan}

\affiliation[b]{Center for Gravitational Physics, Yukawa Institute for Theoretical Physics, Kyoto University, \\
Kyoto 606-8502, Japan}

\affiliation[c]{RIKEN Nishina Center (RNC), \\
Saitama 351-0198, Japan}

\affiliation[d]{Institute of Particle and Nuclear Studies, High Energy Accelerator Research Organization (KEK), \\
Ibaraki 305-0801, Japan}

\affiliation[e]{Research Center for Nuclear Physics (RCNP), Osaka University, \\
Osaka 567-0047, Japan}

\emailAdd{takuya.sugiura@riken.jp}
\emailAdd{yutaro.akahoshi@yukawa.kyoto-u.ac.jp}
\emailAdd{aoym@post.kek.jp}
\emailAdd{takahiro.doi@rcnp.osaka-u.ac.jp}
\emailAdd{doi@ribf.riken.jp}

\abstract{
The nuclear forces are determined by combining 
the HAL QCD method and a new type of source smearing technique.
The new smearing is a projection to a space spanned by
the lowest-lying eigenvectors of the free Laplacian operator on a lattice,
which enables efficient calculation of hadron correlators
at an affordable cost by utilizing the hadron-level momentum conservations.
We find that this new approach reduces the statistical and systematic errors
in the resultant nuclear forces.
}

\FullConference{%
 The 38th International Symposium on Lattice Field Theory, LATTICE2021
  26th-30th July, 2021
  Zoom/Gather@Massachusetts Institute of Technology
}

\begin{document}
\maketitle

\section{Introduction}

Theoretical determination of the nuclear force is a long-standing issue in nuclear physics.
Recent developments of the HAL QCD method 
have made possible determination of the nuclear force from large-scale lattice QCD simulations~\cite{Ishii:2006ec,Aoki:2009ji,Ishii:2012ssm}.
The same method has also been applied to the hyperon forces and 
the 3-body nuclear forces~\cite{Aoki:2012tk}.

Although the nuclear force is the most important target for the HAL QCD method,
its precision determination 
has been difficult, since the nucleon consists of the light quarks only.
To overcome this difficulty, introducing an improved source operator will be important.
In this report, we study the $^1S_0$ and the $^3S_1+^3D_1$ nuclear forces by
combining the HAL QCD method and a new type of smeared source, the free Laplacian Heaviside 
smearing. 
We show below that the new source smearing enables an efficient calculation of the 
nucleon four-point correlators at an affordable cost.

%%%%%%%%%%%%%%%%%%%%%%%%%%%%%%%%%%%%%%%%%%%
\section{HAL QCD method}
%%%%%%%%%%%%%%%%%%%%%%%%%%%%%%%%%%%%%%%%%%%

The equal-time Nambu-Bethe-Salpeter (NBS) wave function for the NN system is defined as
\begin{align}
\psi_{\alpha\beta}(\bm{r};W) 
=
\frac{1}{Z} \braket{0| p_{\alpha}(\bm{r},0) n_{\beta}(\bm{0},0) | 2N;W },
\end{align}
where $\ket{2N;W}$ is the NN asymptotic state with total energy $W=2\sqrt{m_N^2+k^2}$,
$p_{\alpha}(\bm{x})$ and $n_{\beta}(\bm{y})$ are the proton and neutron point-sink operators,
and $Z$ is the wave-function renormalization factor.
In the elastic region $W<W_{\mathrm{th}}$, the asymptotic behavior of the NBS wave function
contains the information of the S-matrix and an energy-independent non-local potential 
$U(\bm{r},\bm{r}^\prime)$ is defined through the Schr\"odinger equation, 
$(E-H_0)\psi_{\alpha\beta}(\bm{r};W) = \int d^3\bm{r}^\prime U_{\alpha\beta;\gamma\delta}(\bm{r},\bm{r}^\prime) \psi_{\gamma\delta}(\bm{r}^\prime)$
with $H_0 \equiv -\Delta/m_N$ and $m_N$ being the nucleon mass~\cite{Ishii:2006ec,Aoki:2009ji}.

The NBS wave function is related to the nucleon four-point correlator as
\begin{align}
\label{eq:R-correlator}
\begin{aligned}
R_{\alpha\beta}(\bm{r},t) 
&\equiv
\frac{1}{Z} 
\braket{0| p_{\alpha}(\bm{r},t) n_{\beta}(\bm{0},t) \overline{\mathcal{J}}(0) | 0 } 
\times e^{2m_N t}
\\
&=
\sum_{n} A_n \psi_{\alpha\beta}(\bm{r};W_n) e^{-\Delta W_n t},
\end{aligned}
\end{align}
where $\overline{\mathcal{J}}(t_0=0)$ is a two-nucleon source operator,
$A_n \equiv \Braket{2N; W_n| \overline{\mathcal{J}}(0) | 0}$ is the overlap factor to the 
$n$-th energy eigenstate in a finite volume, and
$\Delta W_n \equiv W_n - 2m_N$.
When $t$ is large enough, the contribution from states with $\Delta W_n > \Delta W_{\mathrm{th}}$ is 
negligible, and the same non-local potential can be determined through 
the following time-dependent Schr\"odinger-like equation~\cite{Ishii:2012ssm},
\begin{align}
\left( 
\frac{1}{4m_N} \frac{\partial^2}{\partial t^2} 
- \frac{\partial}{\partial t} 
- H_0
\right)
R_{\alpha\beta}(\bm{r},t) 
= 
\int
d^3 \bm{r}^\prime 
U_{\alpha\beta;\gamma\delta}(\bm{r},\bm{r}^\prime) R_{\gamma\delta}(\bm{r}^\prime,t).
\end{align}
In this report we consider the two-nucleon system in the $^1S_0$ and $^3S_1+^3D_1$ channels
and employ the LO approximation of the derivative expansion. 
The potentials $V_C^{(S=0)}(r)$, $V_C^{(S=1)}(r)$ and $V_T^{(S=1)}(r)$ are determined by
solving
\begin{align}
\label{eq:def_1S0_potential}
\left( 
\frac{1}{4m_N} \frac{\partial^2}{\partial t^2} 
- \frac{\partial}{\partial t} 
- H_0
\right) 
\mathcal{P}^{^1S_0} R(\bm{r},t) 
&=
V_C^{(S=0)} (r) \mathcal{P}^{^1S_0} R(\bm{r},t) ,
\\
%%%
\label{eq:def_3S1_3D1_potential}
\begin{bmatrix}
	\left(
	\frac{1}{4m_N} \frac{\partial^2}{\partial t^2} 
	- \frac{\partial}{\partial t} 
	- H_0
	\right) 
	\mathcal{P}^{^3S_1} R(\bm{r},t)
	\\
	\left(
	\frac{1}{4m_N} \frac{\partial^2}{\partial t^2} 
	- \frac{\partial}{\partial t} 
	- H_0a
	\right) 
	\mathcal{P}^{^3D_1} R(\bm{r},t)
\end{bmatrix}
&=
\begin{bmatrix}
	\mathcal{P}^{^3S_1} R(\bm{r},t)		&	\mathcal{P}^{^3S_1} S_{12} R(\bm{r},t)	\\
	\mathcal{P}^{^3D_1} R(\bm{r},t)		&	\mathcal{P}^{^3D_1} S_{12} R(\bm{r},t)	\\
\end{bmatrix}
\begin{bmatrix}
	V_C^{(S=1)}(r) \\
	V_T^{(S=1)}(r)
\end{bmatrix},
\end{align}
where $\mathcal{P}^{^{2S+1}L_J}$ is the projection operator to the $^{2S+1}L_J$ state
and 
$S_{12}\equiv 3(\bm{\sigma}_1 \cdot \hat{\bm{r}})(\bm{\sigma}_2 \cdot \hat{\bm{r}}) - \bm{\sigma}_1 \cdot \bm{\sigma}_2$ 
is the tensor operator.

%%%%%%%%%%%%%%%%%%%%%%%%%%%%%%%%%%%%%%%%%%%
\section{Free LapH Smearing}
%%%%%%%%%%%%%%%%%%%%%%%%%%%%%%%%%%%%%%%%%%%

To arrive at Eqs.~\eqref{eq:def_1S0_potential} and \eqref{eq:def_3S1_3D1_potential},
we have assumed two conditions: 
the elastic-state saturation and the LO dominance of the derivative expansion.
To satisfy the former condition, the temporal separation should be large 
such that $t \gg (W_{\mathrm{th}}-W)^{-1}$,
whereas the signal-to-noise ratio for $R(\bm{r},t)$ tends to
$\mathcal{S}/\mathcal{N} \sim e^{-(m_N - 3 m_\pi/2)t}$ 
and becomes exponentially worse for larger $t$.
Thus, a potential determination should be evaluated at $t$ 
where the systematic errors from the inelastic state contaminations 
and the statistical errors are both controlled.
One way to achieve this is to use an improved source operator such that $A_n$ is small for
all the inelastic states.
In this report, we introduce the free-Laplacian-Heaviside (fLapH) smearing for the 
source operator for this purpose.

The fLapH smearing operator $\mathcal{S}$ is defined as
\begin{align}
\label{eq:LapH_smearing_op}
\mathcal{S}(\bm{x},\bm{y}) = \sum_{l=1}^{N_l} \omega_l v_l(\bm{x}) v_l^*(\bm{y}),
\end{align}
where $v_l(\bm{x})$ are the plain waves, which are $l$-th lowest eigenvectors of the free Laplacian operator,
\begin{align}
\label{eq:free_Laplacian}
\Delta(\bm{x},\bm{y}) = 
\sum_{k=1}^3 
\left\{
\delta(\bm{y},\bm{x}+\hat{\bm{k}}) + \delta(\bm{y},\bm{x}-\hat{\bm{k}}) - 2 \delta(\bm{y},\bm{x})
\right\},
\end{align}
and $\omega_l$ are arbitrary real parameters to control the weight of each eigenmode. 
A similar smearing operator has been introduced in Ref.~\cite{HadronSpectrum:2009krc},
but with the gauge-covariant Laplacian operator instead of Eq.~\eqref{eq:free_Laplacian}.
We will refer to it as covariant LapH.
It is notable that the computational cost of the four-point correlator with the fLapH-smeared source is 
much smaller than that with cLapH-smeared source, as we will see below.

The fLapH smearing is a projection onto a subspace spanned by the $N_l$ lowest-lying 
eigenmodes of the free Laplacian.
The number of eigenmodes $N_l$ will be chosen to be the number of eigenmodes with 
eigenvalues satisfying $-\lambda_l \leq \Lambda^2 \equiv (2\pi |\bm{n}| /L)^2$ with a given $\bm{n}\in \mathbb{Z}^3$.
This means that the fLapH smearing is equivalent to introducing a momentum cutoff $|\bm{p}|\leq \Lambda$.
Since our target is low-energy NN scattering, contribution from $|\bm{p}|> \Lambda$ will be 
small and can be safely neglected.
Also, one can easily see that the wall source corresponds to the $N_l=1$ case of fLapH
and the point source corresponds to $N_l=(L/a)^3$.

The fLapH-smeared NN source operator is defined as 
\begin{align}
\overline{\mathcal{J}}(t_0) 
&=
\overline{p}_{\alpha^\prime}(+\bm{q},t_0) \overline{n}_{\beta^\prime}(-\bm{q},t_0),
\\
%%%%%
\overline{p}_{\alpha^\prime}(+\bm{q},t_0)
&\equiv
+ \epsilon_{abc} \,
\bar{u}^{l_1}_{a \alpha^\prime} (t_0)
\left( \bar{u}^{l_1 T}_{b}(t_0) (C\gamma_5) \bar{d}^{l_2}_{c}(t_0) \right)
\bar{V}_{l_1 l_2 l_3} (+\bm{q}),
\\
%%%%%
\overline{n}_{\beta^\prime}(+\bm{q},t_0)
&\equiv
- \epsilon_{abc} \,
\bar{d}^{l_4}_{a \beta^\prime} (t_0)
\left( \bar{u}^{l_5 T}_{a}(t_0) (C\gamma_5) \bar{d}^{l_6}_{b}(t_0) \right)
\bar{V}_{l_4 l_5 l_6} (-\bm{q}),
\end{align}
where $\pm \bm{q}$ being the relative momentum between the proton and the neutron, and
\begin{align}
\label{eq:smeared_quark_operator}
\bar{q}^{l}_{\alpha c}(t_0) 
&\equiv 
\int d^3 \bm{x} \bar{q}_{\alpha c}(\bm{x},t_0) v^l(\bm{x}),
\\
\label{eq:src_vec_trio}
\bar{V}_{l_1 l_2 l_3} (\pm \bm{q})
&\equiv
\int d^3 \bm{x} \exp\left( \pm i \bm{q}\cdot \bm{x} \right)
\omega_{l_1} \omega_{l_2} \omega_{l_3}
\,
v_{l_1}^{*} (\bm{x}) v_{l_2}^{*} (\bm{x}) v_{l_3}^{*} (\bm{x}).
\end{align}

The quark propagators 
$D_{q}^{-1} v^l = \int d^3\bm{y} \Braket{q(\bm{x},t) \bar{q}(\bm{y},t_0)} \cdot v^l(\bm{y})$
is obtained by solving $Dy = v^l$. 
The calculation of the four-point correlator is performed by the block algorithm~\cite{Doi:2012xd},
where summation over the sink color, spinor, and level indices are first taken as greedily as possible in each diagram, and then the proton block and the neutron block are 
combined to evaluate the correlator.
In the covariant LapH method, the computational cost
of the four-point correlator w.r.t. the color and level indices is 
$\mathcal{O}((N_c N_l)^4)$~\footnote{
With a given cutoff, the number of levels in the covariant LapH method 
is effectively $N_c$ times as large as that of the fLapH, since the former uses the
eigenvectors of the gauge-covariant Laplacian operator.
}.
In the fLapH method, source color indices are independent of the level indices,
resulting in a cost reduction by a factor of $(N_c\cdot N_c!)$ compared to the covariant LapH.
Moreover, one finds that the cost for summation of the level indices is becomes
$\mathcal{O}(N_l^3)$ with the fLapH method.
The latter is because of the momentum conservation relation,
\begin{align}
\label{eq:momentum_conservation}
\bar{V}_{l_1 l_2 l_3} (\pm \bm{q}) = 0
\quad\quad 
\text{if  } \bm{p}_{l_1} + \bm{p}_{l_2} + \bm{p}_{l_3} \neq \pm \bm{q}.
\end{align}
Only the physically allowed combinations of a set of 
source level indices $(l_1,l_2,l_3)$ contribute to each block, 
and the cost of block algorithm is reduced by a factor of $\mathcal{O}(N_l)$
relative to the covariant LapH case.
The momentum conservation relation is not satisfied with covariant LapH, because the 
base vectors are not the momentum eigenstates.
To summarize, the computational cost with the fLapH method is 
$\mathcal{O}(N_l^3)$,
while that with the covariant LapH is $\mathcal{O}(N_l^4)$.
Since calculation of the hadron correlators are generally costly in multi-hadron systems,
fLapH smearing is more promising than covariant LapH in the HAL QCD method.

%%%%%%%%%%%%%%%%%%%%%%%%%%%%%%%%%%%%%%%%%%%
\section{Lattice QCD Setup}
%%%%%%%%%%%%%%%%%%%%%%%%%%%%%%%%%%%%%%%%%%%

We employ the (2+1)-flavor QCD gauge configurations generated by the 
PACS-CS Collaboration~\cite{PACS-CS} on a $32^3\times 64$ lattice 
with the renormalization group improved 
gauge action at $\beta=1.9$ and the non-perturbatively $\mathcal{O}(a)$ improved
Wilson quark action at $c_{SW}=1.715$.
The lattice spacing is $a=0.0907(13)\,\mathrm{fm}$ ($a^-1 \simeq 2176\,\mathrm{MeV}$).
We use the hopping parameters $\kappa_{ud}=0.13700$ and $\kappa_s=0.13640$.
The pion mass is $m_{\pi} \simeq 702\,\mathrm{MeV}$, and the nucleon mass is $m_N \simeq 1581\,\mathrm{MeV}$.
We consider the fLapH smearing combined with the Coulomb gauge fixing and the number of eigenmodes are
$N_l=1, 7, 19$, which correspond to momentum cutoffs $\Lambda$ of
$0\,\mathrm{MeV}$, $2\pi/ L = 427\,\mathrm{MeV}$, and $2\sqrt{2}\pi/ L = 604\,\mathrm{MeV}$, respectively.
We consider three types of weight factors: 
(1) flat : $\omega_l=1$ for all $l$, 
(2) baryon rms optimized : $\omega_l$ is chosen such that
$\sqrt{\Braket{r^2}^b} \equiv \sqrt{\sum_{\bm{r}} \bm{r}^2 |\mathcal{S}(\bm{r},\bm{0})|^6 }$ 
is minimized,
and (3) quark rms optimized : 
similarly $\sqrt{\Braket{r^2}^q} \equiv \sqrt{\sum_{\bm{r}} \bm{r}^2 |\mathcal{S}(\bm{r},\bm{0})|^2 }$ is minimized. The weight factors are summarized in Table.~\ref{table:flaph_parameters}.
To consider the S-wave, the source baryon relative momentum is set to $\bm{q}=0$.

We quark propagators are calculated with Bridge++~\cite{bridge}. 
The periodic (Dirichlet) boundary condition is employed for the spatial (temporal) direction.
The number of measurements is 399 configurations times 2 (for the average of forward/backward propagations in time).

\begin{table}[h]                                                                                                                            
        \centering
        \caption{
        \label{table:flaph_parameters}
        Weight factors for fLapH smearing on a $L/a=32$ lattice
        }
        \begin{tabular}{r|ccc}
        \hline
        $N_l$		&
				$(\omega_{l=1}, \omega_{l=2-7}, \omega_{l=8-19})_{\text{flat}}$	&
					$(\omega_{l=1}, \omega_{l=2-7}, \omega_{l=8-19})_{\text{baryon rms}}$	&
						$(\omega_{l=1}, \omega_{l=2-7}, \omega_{l=8-19})_{\text{quark rms}}$
			\\
        \hline\hline
	   $1$		&
			$(1,0,0)$	&
				-	&
					- 
			\\
	   $7$		&
			$(1,1,0)$	&
				$(0.426194,0.223607,0)$	&
					$(0.724902,0.281223,0)$ 
			\\ 
	   $19$		&
			$(1,1,1)$	&
				$(0.040084,0.121525,0.116206)$	&
					$(0.497497,0.288722,0.145010)$	
			\\
	   \hline
        \end{tabular}
\end{table}

%%%%%%%%%%%%%%%%%%%%%%%%%%%%%%%%%%%%%%%%%%%
\section{Results and discussion}
%%%%%%%%%%%%%%%%%%%%%%%%%%%%%%%%%%%%%%%%%%%

\begin{figure}[tb]
  \centering
  \includegraphics[angle=270,width=0.5\hsize]{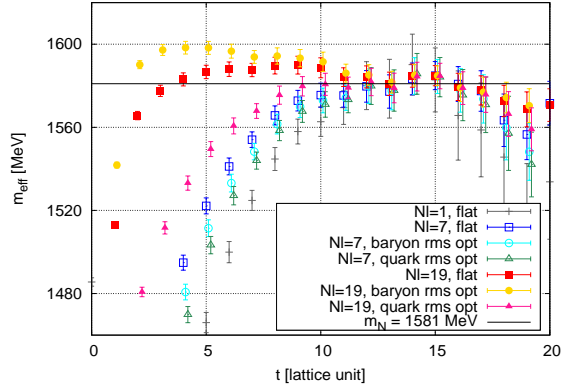}
  \caption{
 	\label{fig:effective_mass}
	Nucleon effective masses
	}
\end{figure}

Shown in Fig.~\ref{fig:effective_mass} are the nucleon effective masses
$m_{\text{eff}}(t) \equiv (1/a)\ln(C(t)/C(t+a))$.
As $N_l$ increases from $1,7,$ to $19$,
we observe a trend in which the plateaux are achieved at smaller $t$
and the statistical errors are drastically reduced.
Also, the effective masses are sensitive to tuning the weight factors $\omega_l$.
There is a visible difference among $N_l=19$ flat (red filled square), 
$N_l=19$ baryon rms opt. (yellow filled circle), 
and $N_l=19$ quark rms opt. (pink filled triangle).
We speculate that this difference is caused by subtle cancellation of 
excited state contributions.

\begin{figure}[t]
  \begin{minipage}{0.5\hsize}
  \centering
  \includegraphics[angle=270,width=\hsize]{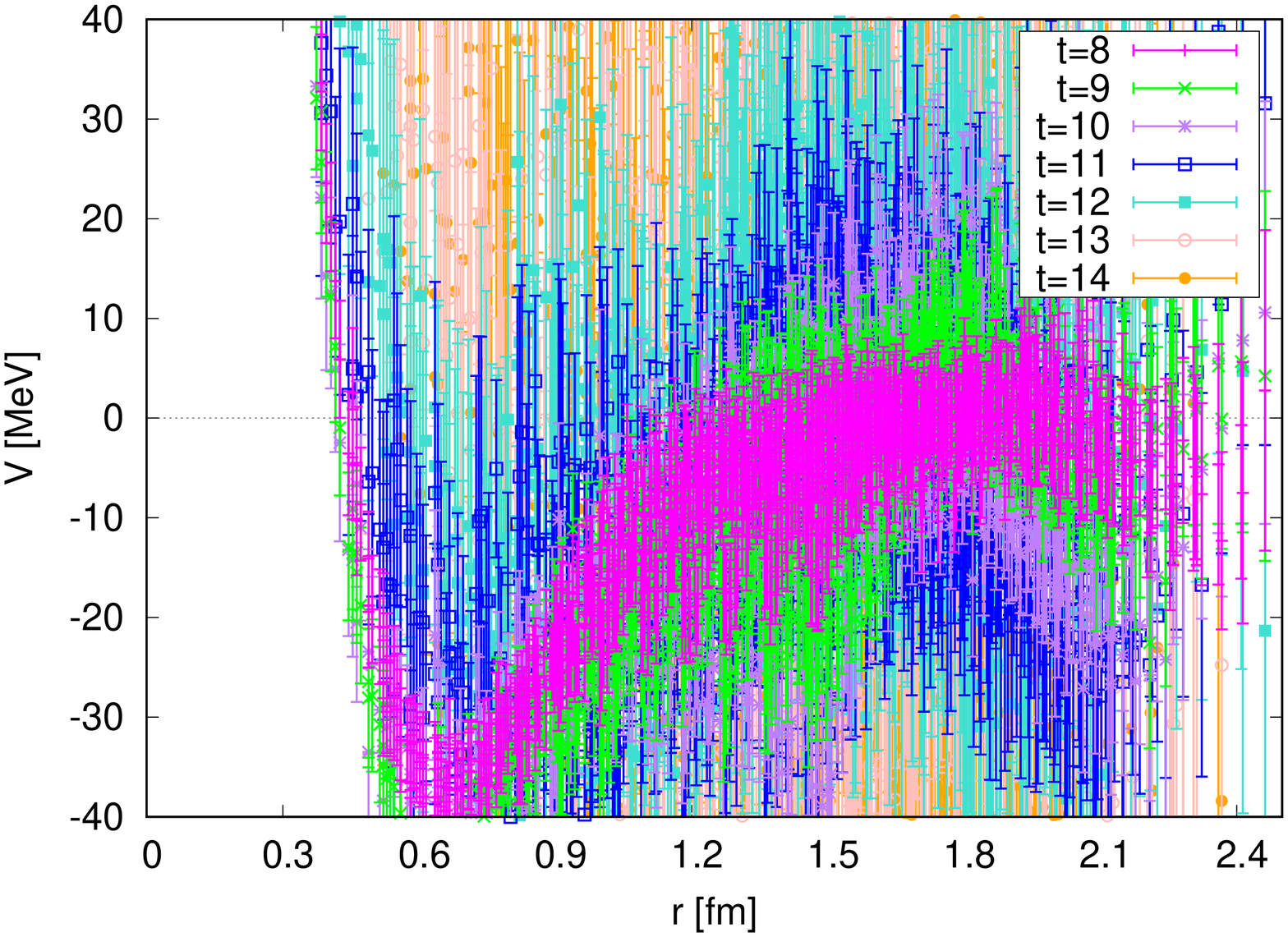}
  \end{minipage}
  \begin{minipage}{0.5\hsize}
  \centering
  \includegraphics[angle=270,width=\hsize]{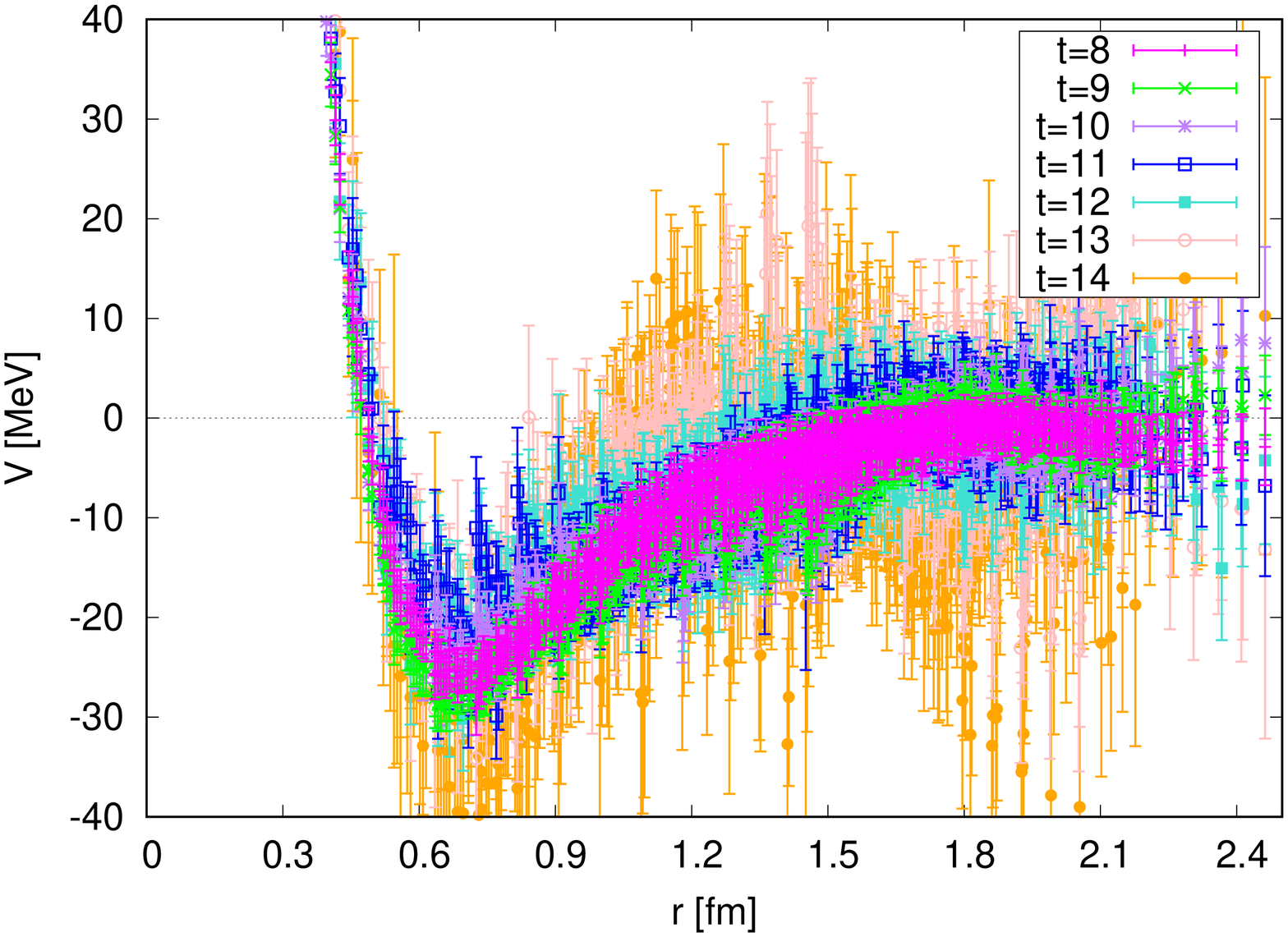}
  \end{minipage}
  \begin{minipage}{0.5\hsize}
  \centering
  \includegraphics[angle=270,width=\hsize]{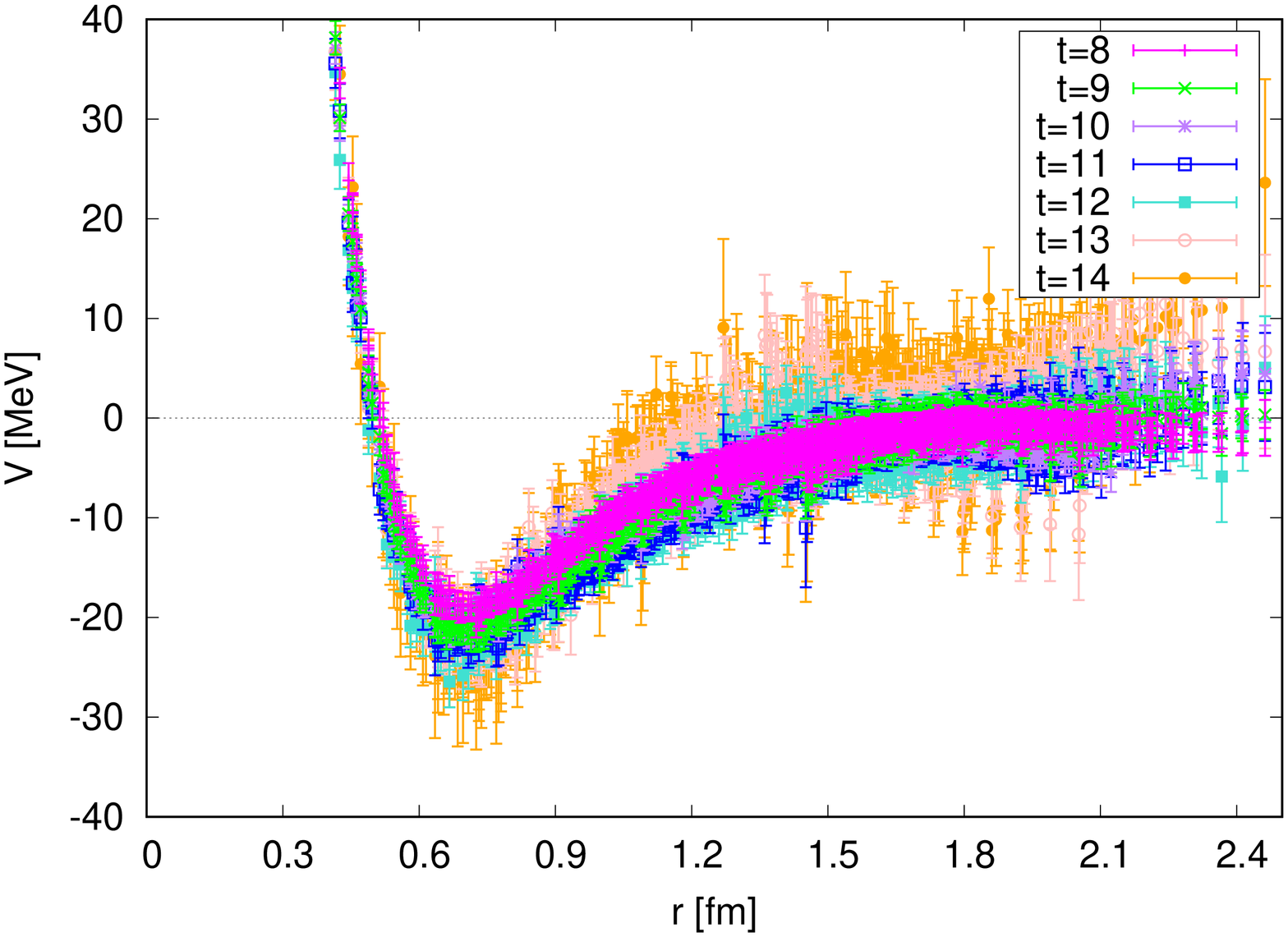}
  \end{minipage}
\caption{
\label{fig:pot_1S0}
The $^1S_0$ nuclear force in Eq.~\eqref{eq:def_1S0_potential} evaluated at $t=8-14$.
(Left Top) $N_l=1$ flat, (Right Top) $N_l=7$ flat, (Left Bottom) $N_l=19$ flat.
}
\end{figure}

\begin{figure}[h]
  \begin{minipage}{0.5\hsize}
  \centering
  \includegraphics[angle=270,width=\hsize]{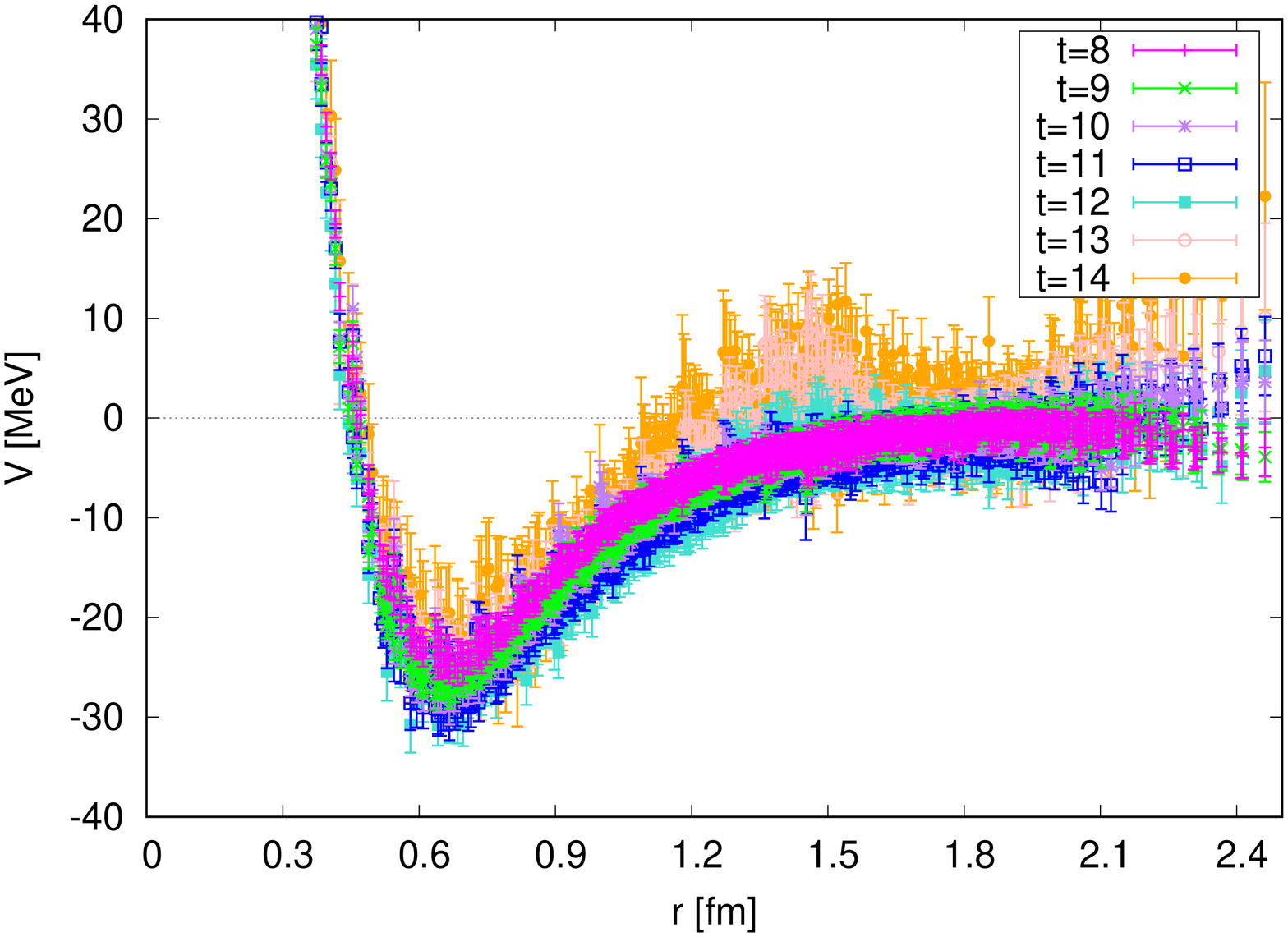}
  \end{minipage}
  \begin{minipage}{0.5\hsize}
  \centering
  \includegraphics[angle=270,width=\hsize]{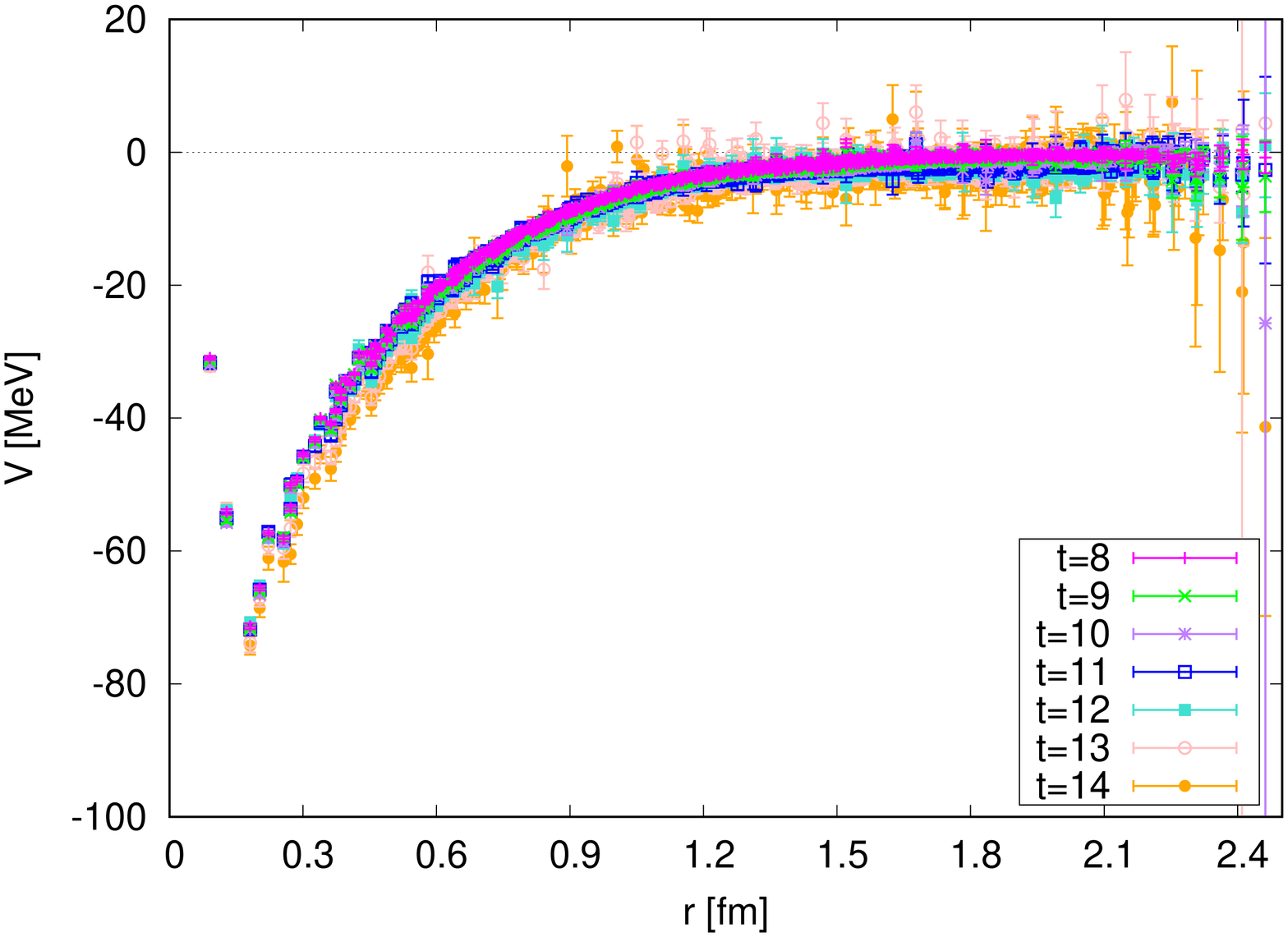}
  \end{minipage}
\caption{
\label{fig:pot_3S1_3D1}
The nuclear force in the $^3S_1+^3D_1$ sector, (Left) the central force $V_C^{(S=1)}(r)$ 
and (Right) the tensor force $V_T^{(S=1)}(r)$ in Eq.~\eqref{eq:def_3S1_3D1_potential}
}
\end{figure}

Figure~\ref{fig:pot_1S0} shows the $^1S_0$ potential $V^{S=0}_C(r)$ in Eq.~\eqref{eq:def_1S0_potential}.
Also, in Fig.~\ref{fig:pot_3S1_3D1}, we show the central force $V_C^{(S=1)}(r)$ and the 
tensor force $V_T^{(S=1)}(r)$ in the $^3S_1+^3D_1$ sector.
We only show the potentials obtained with the flat weights, 
because those with the baryon rms optimized weights or the quark rms optimized weights
show very minor difference from the flat weights case.

The $N_l=1$ case corresponds to the wall source, which has been used in previous 
HAL QCD calculations. 
In fact, our $N_l=1$ result agrees within statistical errors with those in Refs.~\cite{Ishii:2006ec,Ishii:2012ssm}, where larger statistics is achieved
by averaging over multiple source time slice.
With $N_l=7$ and $19$, the statistical errors are drastically reduced compared to the
$N_l=1$ (wall) source.
This reduction of the errors can worth the larger computation cost, as far as
$N_l$ is not too large.
That is especially true on a small lattice, since the required number of modes for
a given momentum cutoff $\Lambda$ tends to $\sim L^3$.
On a large lattice, stochastic estimation of the correlator may be necessary.
Another promising method on a large lattice is the one-end trick~\cite{Akahoshi:2021sxc}.

%%%%%%%%%%%%%%%%%%%%%%%%%%%%%%%%%%%%%%%%%%%
\section{Conclusion}
%%%%%%%%%%%%%%%%%%%%%%%%%%%%%%%%%%%%%%%%%%%

In this report, we have studied the nuclear forces in  the $^1S_0$ and the $^3S_1+^3D_1$ sectors
by using the HAL QCD method and the free Laplacian Heaviside (fLapH) smearing.
The computational cost of the four-point correlators with the fLapH source 
can be drastically reduced compared to those with the commonly-used covariant LapH source,
thanks to the momentum conservation.
The results show that the statistical errors are much smaller than results 
with the wall-source, which have been often used in the HAL QCD calculations.
We conclude that such improvement in the precision is worth the larger computational cost
than the wall source.

We will continue on this study to determine the S-wave nuclear forces with high precision.
Also, the fLapH smearing can be also used for the nuclear force in the P-wave baryon interactions
including the LS force, which is related to a lot of important physics.

%%%%%%%%%%%%%%%%%%%%%%%%%%%%%%%%%%%%%%%%%%%
\section*{Acknowledgements}
%%%%%%%%%%%%%%%%%%%%%%%%%%%%%%%%%%%%%%%%%%%

We thank the members of the HAL QCD Collaboration for fruitful discussion and suggestions.
We thank the PACS-CS Collaboration~\cite{PACS-CS} and ILDG/JLDG~\cite{ijldg} for providing
the gauge configurations.
The lattice QCD calculations have been performed on supercomputer Fugaku at RIKEN.
The calculation of quark propagators has been done by Bridge++ code~\cite{bridge}.
This work is partially supported by HPCI System Research Project 
(hp120281, hp200095, hp200130, hp210165, hp210117),
JSPS Grant (JP18H05236, JP16H03978, JP19K03879, JP18H05407), 
MOST-RIKEN Joint Project ``Ab initio investigation in nuclear physics'', 
``Priority Issue on Post-K computer'' (Elucidation of the Fundamental Laws and
Evolution of the Universe), 
``Program for Promoting Researches on the Supercomputer Fugaku''
(Simulation for basic science: from fundamental laws of particles to creation of nuclei) and Joint
Institute for Computational Fundamental Science (JICFuS).

%%%%%%%%%%%%%%%%%%%%%%%%%%%%%%%%%%%%%%%%%%%

\end{document}